\documentclass[11pt,a4paper]{article}

\usepackage{jheppub}
\usepackage{enumerate}
\usepackage{graphicx}
\usepackage{bm}
\usepackage{dcolumn}
\usepackage{color}
\usepackage{pst-plot}

\newcommand{\ave}[1]{\left\langle #1 \right\rangle}
\newcommand{\order}[1]{ \mathcal{O} \left( #1 \right) }

\renewcommand{\texttt}{{}}
\newcommand{\be}{\begin{eqnarray}}
\newcommand{\ee}{\end{eqnarray}}

\title {The Hawking-Page crossover in noncommutative anti-deSitter space}
\author[a,b]{Piero Nicolini}
\author[a,b]{Giorgio Torrieri}
\affiliation[a]{Frankfurt Institute for Advanced Studies (FIAS), 
Frankfurt am Main, Germany}
\affiliation[b]{
Institut f\"ur Theoretische Physik,  J. W.
Goethe-Universit\"at,
Frankfurt am Main, Germany}

\emailAdd{nicolini@th.physik.uni-frankfurt.de}
\emailAdd{torrieri@th.physik.uni-frankfurt.de}

\abstract{
We study the problem of a Schwarzschild-anti-deSitter black hole in a noncommutative geometry framework, thought to be an  effective description of quantum-gravitational spacetime. As a first step we derive the noncommutative geometry inspired Schwarzschild-anti-deSitter solution. After studying the horizon structure, we find that the curvature singularity is smeared out by the noncommutative fluctuations.  On the thermodynamics side, we show that the black hole temperature, instead of a divergent behavior at small scales, admits a maximum value. This fact implies an extension of the Hawking-Page transition into a van der Waals-like phase diagram, with a critical point at a critical cosmological constant size in Plank units and a smooth crossover thereafter.   We speculate that, in the gauge-string dictionary, this corresponds to the confinement ``critical point'' in number of colors at finite number of flavors, a highly non-trivial parameter that can be determined through lattice simulations.
}

\begin{document}

\maketitle
\flushbottom

\section{Introduction}

Since the inception of the old program of semiclassical quantum gravity, there was evidence about the possibility of interpreting a black hole as a thermodynamic system in which the role of the temperature was played by the surface gravity, i.e,  the strength of the gravitational field at the event horizon \cite{Haw74}. This has been possible since thermodynamics permits a description of a system in terms of macroscopic variables even if we ignore all the details of the underlying microscopic statistical theory. In other words, we can study black hole thermodynamics by merely consider general relativity as a low energy limit of a yet not completely known microscopic description of gravity, i.e,  quantum gravity. This is true even when we work in the framework of ``Gauge-string correspondence'', a duality between gravitational physics in anti-deSitter space and particular kinds of gauge field theory in one dimension fewer \cite{ads1,ads2,ads3}. In such a case, there is an advantage: the anti-deSitter spacetime acts like a box with reflecting walls, which tame the usual thermodynamic instability of black holes in asymptotically flat space \cite{HaP83}.  Despite many efforts and remarkable results in specific cases \cite{StV96,CaM96}, we currently cannot claim to having a satisfactory description of the general properties of black hole thermodynamics in the extreme energy regime, namely when some ultraviolet completion of gravity must be invoked.

In the last years, there have been attempts to circumvent this problem by following an alternative route: instead of adopting a full quantum gravity formulation, one can think to focus on the specific issue of black hole singularities and provide a description of the microscopic statistical nature of gravity by an effective theory.
Examples of these attempts have been inspired by a variety of formulations, e.g.,  noncommutative geometry, loop quantum gravity, generalized uncertainty principles, nonlocal gravity and asymptotically safe gravity.  In most of these cases, the coarse grained structure of spacetime in its extreme regime naturally provides an effective ultraviolet cutoff \cite{BoR00,Mod04,Nic05,MMN11}.  As a result families of singularity free black hole geometries have been derived. Even if these black holes are due to different formulations, they seem to agree about a stable thermodynamic behaviour: in the majority of models, including the neutral, non rotating case, the black hole temperature admits a maximum which marks a change from a negative to a positive heat capacity configuration. In addition, there is a prevalence of cases in which black holes do not evaporate off, but cool down into a zero temperature extremal black hole remnant. This fact is having potential repercussions about our understanding of the quantum stability of the deSitter space: within the instanton formalism it has been shown that, contrary to expectations, the production of Planck size black holes would have been suppressed in inflationary epochs of the universe \cite{MaN11}.

In this paper we want to extend this recent analysis of quantum gravity improved black holes in deSitter space to the case of anti-deSitter background. This is not a simple exercise of sign change in the cosmological constant, but a study about how the new black hole thermodynamics combines with the anti-deSitter ``box'' with reflecting walls. This is relevant beyond mere considerations about black hole thermodynamics: the possibility of horizon extremisation even in the neutral case may disclose a new phase structure for the transition between the confined and the deconfined regime of the holographycally dual field theory, in view of the relationship between the Hawking-Page phase transition and confinement via the gauge-string correspondence \cite{adswitten,adswitten2}

As in the asymptotically deSitter case \cite{MaN11}, we follow the strategy of employing noncommutative geometry inspired black holes as models of quantum gravity black holes. This family of metrics has been determined thanks to the possibility of smearing out curvature singularities by considering the average effect of the fluctuations of a noncommutative manifold \cite{Nic09}. The paper is organized as follows: in section \ref{line}, after recalling some basic properties of noncommutative inspired black holes, we derive the the noncommutative geometry inspired Schwarzschild-anti-deSitter (NCSchwAdS) and we study its thermodynamics; in section \ref{gauge} we draw a scenario about the consequences for the dual field theory; in section \ref{concl} we summarize the results  and we draw the conclusions.

\section{The noncommutative inspired Schwarzschild-anti-deSitter spacetime \label{line}}

\subsection{The line element}

To determine the NCSchwAdS line element we can proceed along the lines of what recently found for the NCSchwdS case in \cite{MaN11}. The basic ingredient behind  noncommutative geometry inspired solutions is the absence of any point like object or better the impossibility of resolving objects to better accuracy than a minimal length.
This idea, at the basis of any formulation of quantum gravity can be realized in a simple and effective way by considering the mean position of an object when coordinate operators fail to commute. In a series of papers \cite{SmS03a,SmS03b,SmS04,SSN06,BGM10} it has been shown that  the average of coordinate operators on suitable coordinate coherent states provides a mean position which is no longer governed by a Dirac delta function but by a Gaussian distribution
\begin{equation}
f_{\ell}(\vec{x})=\frac{1}{(4\pi\ell^2)^{d/2}} e^{-|\vec{x}|^2/4\ell^2},
\label{gaussian}
\end{equation}
where $d$ is the dimension of a Euclidean manifold and $\ell$, is the minimal length implemented through the noncommutative relations among coordinate operators. In the absence of extradimensions, a natural choice for $\ell$ would be a value of the order of the Planck length, i.e., $\ell\sim\sqrt{G}\approx 1.6\times 10^{-33}$ cm.
However the value of $\ell$ is not fixed \textit{a priori} by the theory and therefore we can treat it as a parameter adjustable to the relevant scale at which noncommutative effects set it.
It has been shown that primary corrections to any field equation in the presence of a noncommutative background can be obtained by replacing the conventional point like source term (matter sector) with a Gaussian distribution, while keeping formally unchanged differential operators (geometry sector) \cite{Nic05,NSS06a,Nic09,BGM10}. In the specific case of the gravity field equations this is equivalent to saying that the only modification occurs at the level of the energy-momentum tensor,  while $G_{\mu\nu}$ is formally left
unchanged.

For a static, spherically symmetric, noncommutative diffused, particle-like gravitational source of mass $M$, one gets a Gaussian profile for the $T_0^0$ component of the  energy-momentum tensor
 \begin{equation}
 T_0^0=-\rho_\ell (r)=-\dfrac{M}{\left(4\pi\ell^2\right)^{3/2}}\,\exp\left(-\frac{r^2}{4\ell^2}\right).
 \label{tgaussian}
\end{equation}
This specific smearing of source terms discloses model independent characters which emerge also in the case of loop quantum gravity \cite{Mod04,Mod06,Mod08,Mod10} and asymptotically safe gravity \cite{BoR99,BoR00,BoR06}, at least at a qualitative level. On the other hand, the model independence issue has already been addressed more quantitatively by observing that the Gaussian profile (\ref{gaussian}) can also be obtained in another context through the action of a nonlocal operator $e^{\ell^2\Delta_x}$
\begin{equation}
f_{\ell}(\vec{x})=e^{\ell^2\Delta_x}\delta(\vec{x}).
\end{equation}
By means of this relation, it has recently been shown that Einstein gravity coupled to a smeared source term like
(\ref{tgaussian}) is dual to an ultraviolet complete nonlocal gravity action coupled to an ordinary source term \cite{Mof10,MMN11}.  As a consequence the formulation preserves its intrinsic nonlocal character also at the level of gravity field equations.  The short distance quantum gravity modifications permits to avoid the occurrence of any curvature singularity by the presence of quantum fluctuations whose average effect is described by a local deSitter geometry. This is the key result at the basis of the whole family of noncommutative geometry inspired black holes \cite{NSS06b,Riz06,ANS07,CaN08,SSN09,ABN09,ABN10,NiS10,SmS10,Gin10,MoN10b,MaN11,MuN11} (and their companion geometries \cite{GaL09,OhP10}), which depart from what happens with the deformation of products of vielbein fields in terms of Moyal $\star$-products. In such a case to obtain the $\star$-deformed equations, one is fatally compelled to expand the Moyal product in the noncommutative parameter, depriving the theory of its nonlocal character \cite{Mof00}.  

The covariant conservation law $\nabla_\mu T^{\mu\nu}=0$ and the ``Schwarzschild like'' condition $g_{00}=-g_{rr}^{-1}$ completely specify the energy momentum tensor, whose form is given by
\begin{equation}
 T^\mu{}_\nu=\mathrm{Diag}\left(\, -\rho_\ell\left(\, r\,\right),\ p_r\left(\,r\,\right),\
 p_\perp\left(\, r\,\right),\ p_\perp\left(\, r\,\right)\,\right).
 \label{stresst}
 \end{equation}
We notice that there are nonvanishing pressure terms with $p_r\neq p_\perp$, corresponding to the case of an anisotropic fluid. Contrary to the conventional picture of matter squeezed at the origin,  here the noncommutative geometry is effectively described as a fluid diffused around the origin. If one substitutes the above energy momentum tensor in Einstein equations one obtains the noncommutative geometry inspired Schwarzschild solution (NCSchw) \cite{NSS06b}, which matches the conventional Schwarzschild spacetime, except at short distances where the curvature singularity is smeared out.

In this paper we want to determine black hole solutions in the presence of a negative cosmological term $\Lambda=-3/L^2$, where $L$ is a parameter with the dimension of a length. We consider the Einstein equations
\be
R_{\mu\nu}-\frac{1}{2}Rg_{\mu\nu}+\Lambda g_{\mu\nu}=8\pi G T_{\mu\nu}
\ee
and a line element of the form
\be
 &&ds^2 = -V(r)\, dt^2 + V(r)^{-1}\, dr^2 +
r^2\,d\Omega^2. 
\ee
In \cite{MaN11} it has been shown that the above equations can be derived by a functional variation of the action
\begin{equation}
I=\int d^4 x\sqrt{-g}\left[\frac{1}{16\pi G}\left(R-2\Lambda \right)+L_m\right]
\end{equation}
with respect to the metric where $L_m=2p_r-p_\perp$.
It is convenient to introduce the following tensor
\be
 {\cal T}_{\mu}^{\nu}\equiv T_{\mu}^{\nu}-\frac{\Lambda}{8\pi G}\delta_{\mu}^{\nu}= \mathrm{Diag}\left(\, {\cal E} \left(\,r\,\right),\ {\cal P}_r\left(\,r\,\right),\
 {\cal P}_\perp\left(\, r\,\right),\ {\cal P}_\perp\left(\, r\,\right)\,\right) \nonumber
\ee
with ${\cal E}(r)=-\rho-\Lambda/8\pi G$,
which yields the following ``fluid'' equations 
\begin{eqnarray}
&&\frac{d{\cal M}}{dr}= 4\pi\, r^2{\cal E}(r)\ ,\label{eq1}\\
&&\frac{1}{2g_{00}}\frac{dg_{00}}{dr}=\frac{{\cal M}(r)+4\pi \ r^3 {\cal P}_r(r)}{r(r-2{\cal M}(r))}\ , \label{eq22}\\
&&\frac{d{\cal P}_r}{dr}= - \frac{1}{2g_{00}}\frac{dg_{00}}{dr}\left(\, {\cal E}+{\cal P}_r\,\right)+\frac{2}{r}\left(\, {\cal P}_\perp-{\cal P}_r \,\right)
\label{eq33}
\end{eqnarray}
We recall that the condition $g_{00}=-g_{rr}^{-1}=-V(r)$ is equivalent to the equation of state
\begin{equation}
 {\cal P}_r(r)=-{\cal E}(r).
\end{equation}
As a result we obtain the NCSchwAdS line element 
\be
V(r)= 1- \frac{4M G \gamma(3/2; r^2/4\ell^2)}{r\sqrt{\pi}}+\frac{ r^2}{L^2}\, \, 
\label{lineel}
\ee
where
\be
\gamma(3/2; x)=\int_0^x dtt^{1/2}e^{-t}.
\ee
The angular pressure turns out to be
\be
p_\perp (r)=-\rho_\ell(r)\left(1-\frac{r^2}{4\ell^2}\right).
\label{pperp}
\ee
We start the analysis of (\ref{lineel}), by noticing that for $r\gg \ell$ the solution coincides  with the conventional Schwarzschild-anti-deSitter (SchwAdS) line element \cite{HaP83}. In other words this is the regime where noncommutative fluctuations are negligible and the spacetime can well described by a classical manifold.
On the other hand, at small length scales, i.e., high energies there is a crucial departure from the conventional scenario. Expanding ($\ref{lineel}$) for $r\ll\ell$ we get
\begin{equation}
V(r)\approx 1-\frac{\Lambda_{\mathrm{eff}}}{3} r^2
\end{equation}
where 
\begin{equation}
\Lambda_{\mathrm{eff}}=-\frac{3}{L^2}+\frac{1}{\sqrt{\pi}}\frac{MG}{\ell^3}.
\end{equation}
The metric is regular at the origin and endowed with an effective cosmological constant $\Lambda_{\mathrm{eff}}$. The term $\frac{1}{\sqrt{\pi}}\frac{MG}{\ell^3}$ is positive and is the effect of the local fluctuations of the geometry which cure the curvature singularity, while $-\frac{3}{L^2}$ is simply the background anti-deSitter term.
According to the value of $M$, we have that 
\begin{enumerate}[i)]
\item for $M>\frac{3}{L^2}\frac{\sqrt{\pi}\ell^3}{G}$ there is a deSitter core at the origin, i.e., $\Lambda_{\mathrm{eff}}>0$ with local gravitational repulsion;
\item for $M<\frac{3}{L^2}\frac{\sqrt{\pi}\ell^3}{G}$ there is a anti-deSitter core at the origin, i.e., $\Lambda_{\mathrm{eff}}<0$ with local gravitational attraction;
\item for $M=\frac{3}{L^2}\frac{\sqrt{\pi}\ell^3}{G}$ there is a local Minkowski space at the origin, i.e., $\Lambda_{\mathrm{eff}}=0$ with no gravitational interaction.
\end{enumerate}
In any of the above cases, the spacetime, for its regularity is geodesically complete. As shown in \cite{MaN11}, if one considers the solution for negative $r$ we do not obtain a mere analytical continuation of the positive $r$ solution, but an additional distinct spacetime. We can write this line element in an equivalent and more correct way by assuming negative values of the mass parameter $M=-|M|$, i.e.,  
\be
V_-(r)= 1+\frac{4|M| G \gamma(3/2; r^2/4\ell^2)}{r\sqrt{\pi}}+\frac{ r^2}{L^2},\, \, 
\ee
which describes a spacetime without horizons and with a negative $\Lambda_{\mathrm{eff}}$.

\begin{figure}
 \begin{center}
 \includegraphics[height=5.5cm]{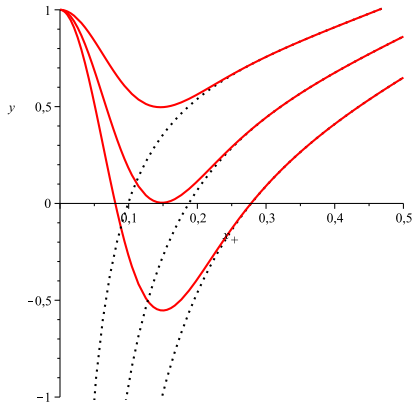}
       \caption{\label{figure0} In figure above we set $q=0.1$. The solid curves from top are the functions $V(x)$ for $m=0.05$, $m=m_0\approx 0.097$ and $m=0.15$. The dotted curves are the corresponding curves for the conventional solution, i.e., $q=0$.  We notice that there exist three cases, namely two horizons, no horizon and one single degenerate horizon.}
     \end{center}
  \end{figure}

An interesting feature of the solution (\ref{lineel}) is the horizon equation $V(r_H)=0$. This equation depends on two parameters, $M$ and $L$ and cannot be solved in a closed form. However we can draw plots and study the occurrence of horizons (see Fig. \ref{figure0}).
To this purpose it is more convenient to write (\ref{lineel}) as
\be
V(x)= 1- \frac{4m  \gamma(3/2; x^2/q^2)}{x\sqrt{\pi}}+x^2\, \, 
\ee
where $x=r/L$, $m=MG/L$ and $q=2\ell/L$, to study plots in $\ell$ units. The parameter $q$ is crucial. It is responsible of a modified causal structure of the solution which resembles the Reissner-Norstr\"{o}m-anti-deSitter case \cite{CEJ99a,CEJ99b}. For this reason we shall call $q$ ``fictitious charge''.  We see that the curve starts at the origin at $V(0)=1$. Close to the origin the curve has quadratic corrections
\be
V(x)= 1- \left(\frac{8m}{3\sqrt{\pi}}q^{-3}-1\right)x^2+{\cal O}(x^2)\, \, 
\ee
For $m\leq \frac{3\sqrt{\pi}}{8}q^3$, the curve keeps growing and never reaches the $x$-axis. This is equivalent to say that in case of anti-deSitter (Minkowski) core at the origin, the spacetime admits no horizon. For masses $m>\frac{3\sqrt{\pi}}{8}q^3$, the curve starts decreasing for small values of $x$, before growing up again at large values.
In such a way the curve can or cannot reach the $x$-axis, depending on the value of $m$. We shall distinguish  three possibilities
\begin{enumerate}[i)]
\item for $m>m_0\geq \frac{3\sqrt{\pi}}{8}q^3$ (and therefore for $M>M_0$), with $m_0=m_0(q)$ ($M_0=M_0(\ell, L)$) there are two horizons, i.e.,  $x_-$ and $x_+$($r_-$ and $r_+$);
\item for $m=m_0>\frac{3\sqrt{\pi}}{8}q^3$ ($M=M_0$) there is one degenerate horizon $x_0$ ($r_0$);
\item for $\frac{3\sqrt{\pi}}{8}q^3<m<m_0$ ($M<M_0$) there is no horizon.
\end{enumerate}
At first sight the above three possibilities look like what happens for the NCSchw solution. However for the NCSchwAdS the threshold mass $M_0$ depends not only on $\ell$, but also on the cosmological parameter $L$.
We can determine $m_0$ by requiring
\be 
V(x)=\frac{\partial V}{\partial x}=0,
\ee
a condition which gives the following system of equations
\be 
\left\{
\begin{array}
[c]{l}%
 m=\frac{\sqrt{\pi}}{4}\frac{x(1+x^2)}{\gamma(3/2; x^2/q^2)}\\
\\
 m=\frac{{x}^{3}
\sqrt {\pi }}{\left(\frac{2 x}{q} +\frac{4{x}^{3}}{q^3} \right){{\rm e}^{-\frac{{x}^{2}}{{q}^{2}}}}-\sqrt {\pi }{{\rm erf}\left(\frac{x}{q}\right)}}.
\end{array}
\right.
\ee
We can numerically solve this system to obtain the values of $m_0$ and $x_0$ for given $q$ (see Tab. \ref{table}).
We recall that the NCSchwdS can admit even three distinct horizons, two related to the black hole and a cosmological horizon \cite{MaN11}.

\begin{table}[h]
\begin{center}
{
\begin{tabular}{ccccccc} 
\hline
 $q \ \ $
 & $1/15$ &  $1/10$ &  $1/6$ &  $q^\ast$ &    $1/3$ &  $1$  \\
\hline
\hline 
 $m_0 \ \ $ 
 &0.064   & 0.097  & 0.168  & 0.186  & 0.393
  &  2.689  \\
\hline
 $x_0 \ \ $   & $0.10$   & $0.15$  & $0.24$  & $0.27$  & $0.40$  &
$0.70$    \\
\hline
\end{tabular} }
\end{center}
\caption{Remnant masses and radii for different values of $q$. }
 \label{table}
\end{table}

\subsection{Thermodynamics}

Before starting the study of the thermodynamics of the NCSchwAdS metric, we briefly recall some properties of the conventional SchwAdS metric , i.e., 
\be && ds^2=  -{\cal V}dt^2+{\cal V}^{-1}dr^2+r^2d\Omega^2   \\
&& {\cal V}= 1-\frac{2MG}{r}+\frac{r^2}{L^2}.
\ee
To compute the horizon temperature, we can make the substitution $\tau=it$ and get a positive defined metric for $r>r_+$. As in the asymptotically flat space, the apparent singularity at $r=r_+$ can be removed if $\tau$ is regarded as an angular coordinate with period
\be  \beta=\frac{4\pi L^2 r_+}{L^2+3r_+^2}, \ee
which is nothing but the inverse of the temperature associated to the horizon $r_+$. Alternatively one can compute the internal energy of the system
\be M={\cal U}(r_+)=\frac{r_+}{2G}+\frac{r_+^3}{2GL^2} \ee
and get the temperature by
\be 
{\cal T}&\equiv &\frac{1}{4\pi}\left. \frac{d{\cal V}(r)}{dr} \right|_{r_+}\nonumber \\ &=&\beta^{-1}=\nonumber \\
&=&\frac{1}{4\pi r_+}\left(1+\frac{3r^2_+}{L^2}\right).
\label{convtemp}
\ee
The above temperature has a minimum value ${\cal T}_{\mathrm{min}}=\sqrt{3}/(2\pi L)$ when $r_+=r_{\mathrm{min}}=L/\sqrt{3}= 1/\sqrt{-\Lambda}$.
The entropy $d{\cal S}\equiv d{\cal U}/{\cal T}$ can be derived from
\be d{\cal S}=\frac{1}{{\cal T}}\frac{\partial {\cal U}}{\partial r_+}dr_+ \ee
which confirms the same relation with the area of the event horizon as in the asymptotically flat case
\be d{\cal S}=\frac{1}{G}\ 2\pi r_+dr_+.  \ee
The heat capacity ${\cal C}\equiv d{\cal U}/dr_+$ can be written as
\be  {\cal C}={\cal T}\left(\frac{d{\cal S}}{dr_+}\right)\left(\frac{d{\cal T}}{dr_+}\right)^{-1}.\ee
At $r_+=r_{\mathrm{min}}$ the quantity $d{\cal T}/dr_+$ vanishes and the heat capacity has a pole, which marks a change of sign
\be {\cal C}=\frac{2\pi r_+^2}{G}
\left(\frac{r_+^2+r_{\mathrm{min}}^2  }{r_+^2 - r_{\mathrm{min}}^2}\right).
 \ee
This means that at $r_+=r_{\mathrm{min}}$ the black hole switches from a (locally) stable configuration at $r_+>r_{\mathrm{min}}$ (${\cal C}>0$) to an unstable configuration at $r_+<r_{\mathrm{min}}$ (${\cal C}<0$).
Thus for a temperature ${\cal T}>{\cal T}_{\mathrm{min}}$ there are two possible horizon radii $r_1<r_{\mathrm{min}}<r_2$. The smaller radius case $r_1$ is therefore unstable to decay either into pure thermal radiation or to the larger value of the horizon radius $r_2$. The free energy of the system is given by
\be {\cal F}\equiv {\cal U}-{\cal T}{\cal S} \ee
and therefore it turns out to be
\be {\cal F}=\frac{r_+}{4G}\left(1-\frac{r_+^2}{L^2}\right). \ee
This means that at $r_1$ the free energy is positive and therefore the black hole configuration is less probable than a pure  thermal  background. The large value of the horizon $r_2$, even if locally stable, can have positive free energy if $r_2<L$, i.e., $T<(\pi L)^{-1}$. Therefore the configuration would reduce its free energy by Hawking emission and eventually evaporate off. This is the so called Hawking-Page phase transition \cite{HaP83}. On the contrary for $r_2>L$, i.e., $T<(\pi L)^{-1}$, the free energy becomes negative and the pure thermal radiation will be disfavoured with respect to the (globally stable) black hole configuration.


\begin{figure}
 \begin{center}
 \includegraphics[height=5.5cm]{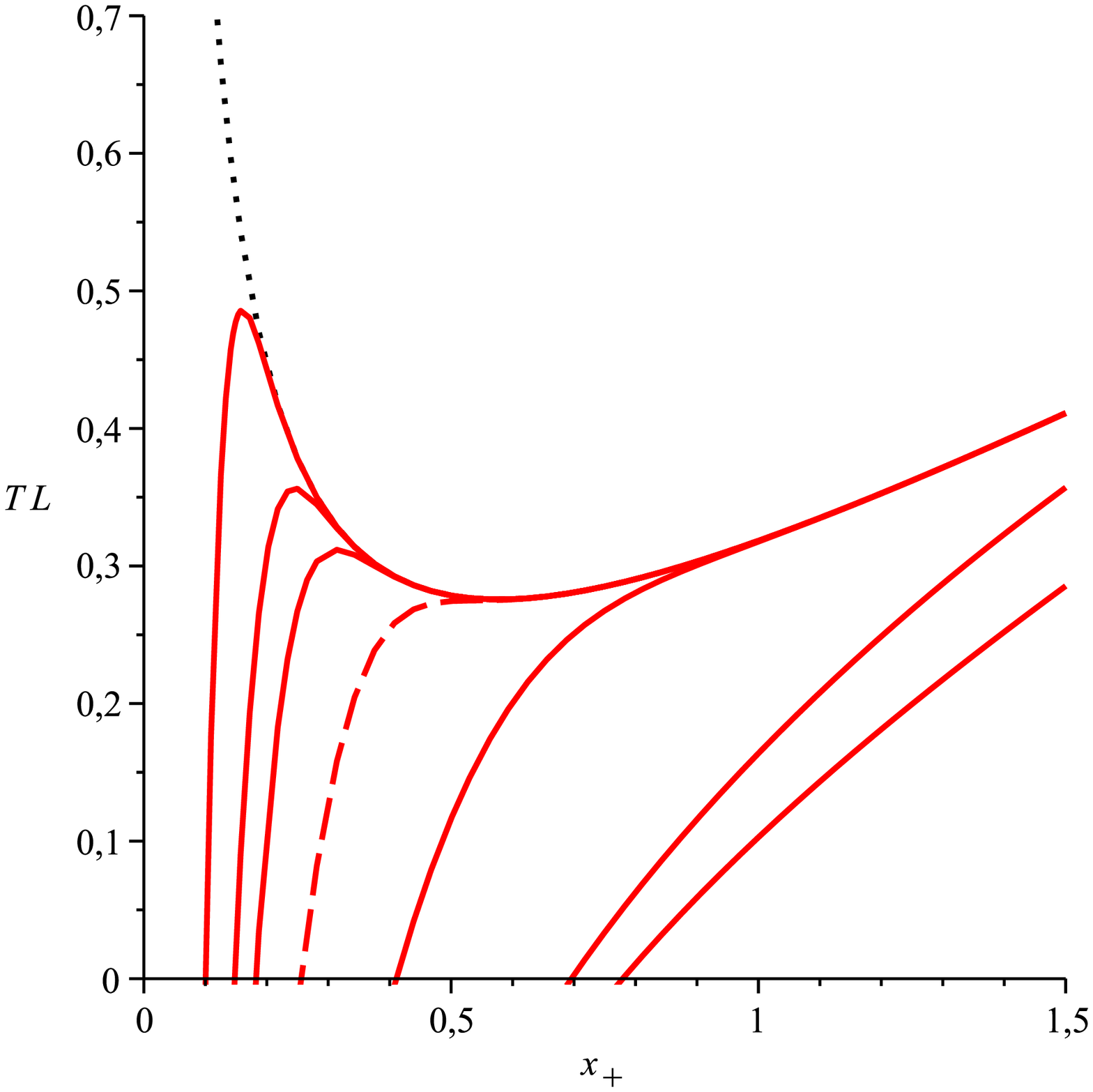}
  \includegraphics[height=5.5cm]{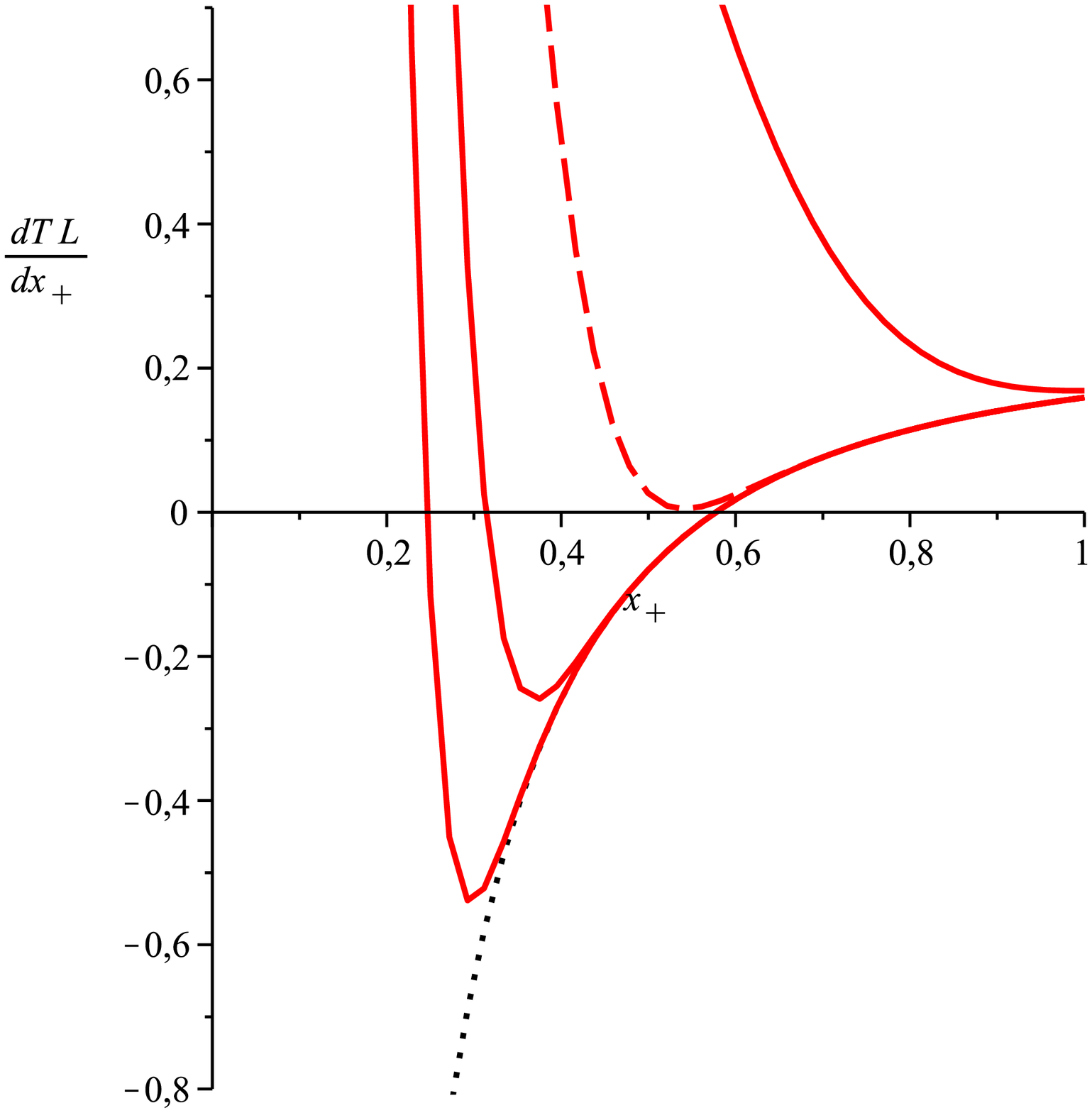}

      \caption{\label{figure1} On the left we have the Hawking temperature $T\times L$ as a function of $x_+$. The dotted curve is the Hawking-Page temperature ($q=0$) and the solid lines, from top to bottom, correspond to the NCSchwAdS temperature for $q=1/15$,  $q=0.1$, $q=1/8$,  $q=1/3$ and $q=1$, $q=2$. The dashed curve corresponds to the critical value $q=q^\ast\approx 0.18243$. On the right we have $L\times dT/dx_+$ as a function of $x_+$. The dotted curve is the Hawking page case ($q=0$). The solid lines from bottom to top correspond to  the NCSchwAdS case for  $q=0.1$, $q=1/8$, $q=1/3$ and again the dashed curve corresponds to the critical value $q=q^\ast\approx 0.18243$.}
     \end{center}
  \end{figure}

  \begin{figure}
 \begin{center}
 \includegraphics[height=5.5cm]{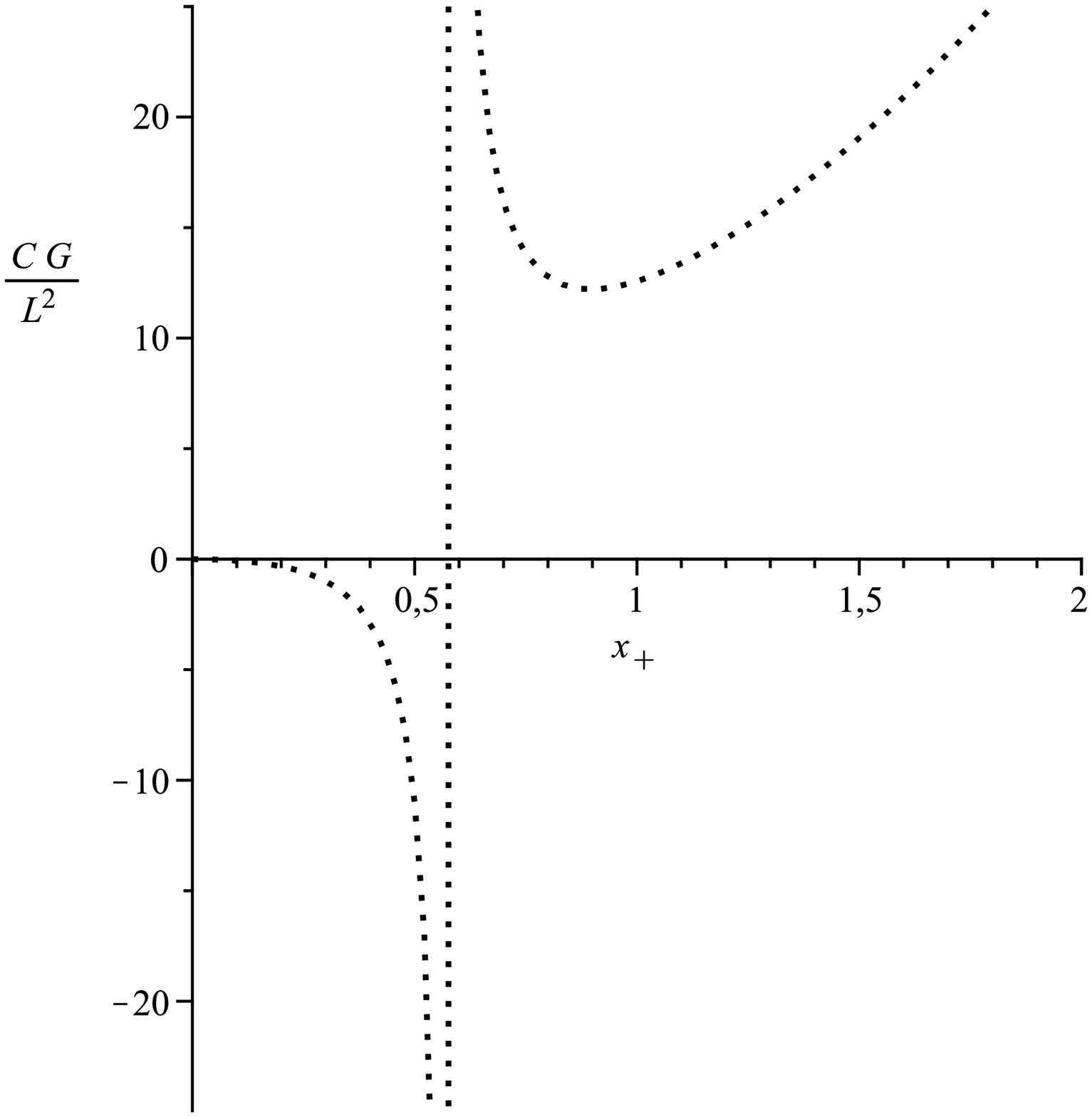}
 \includegraphics[height=5.5cm]{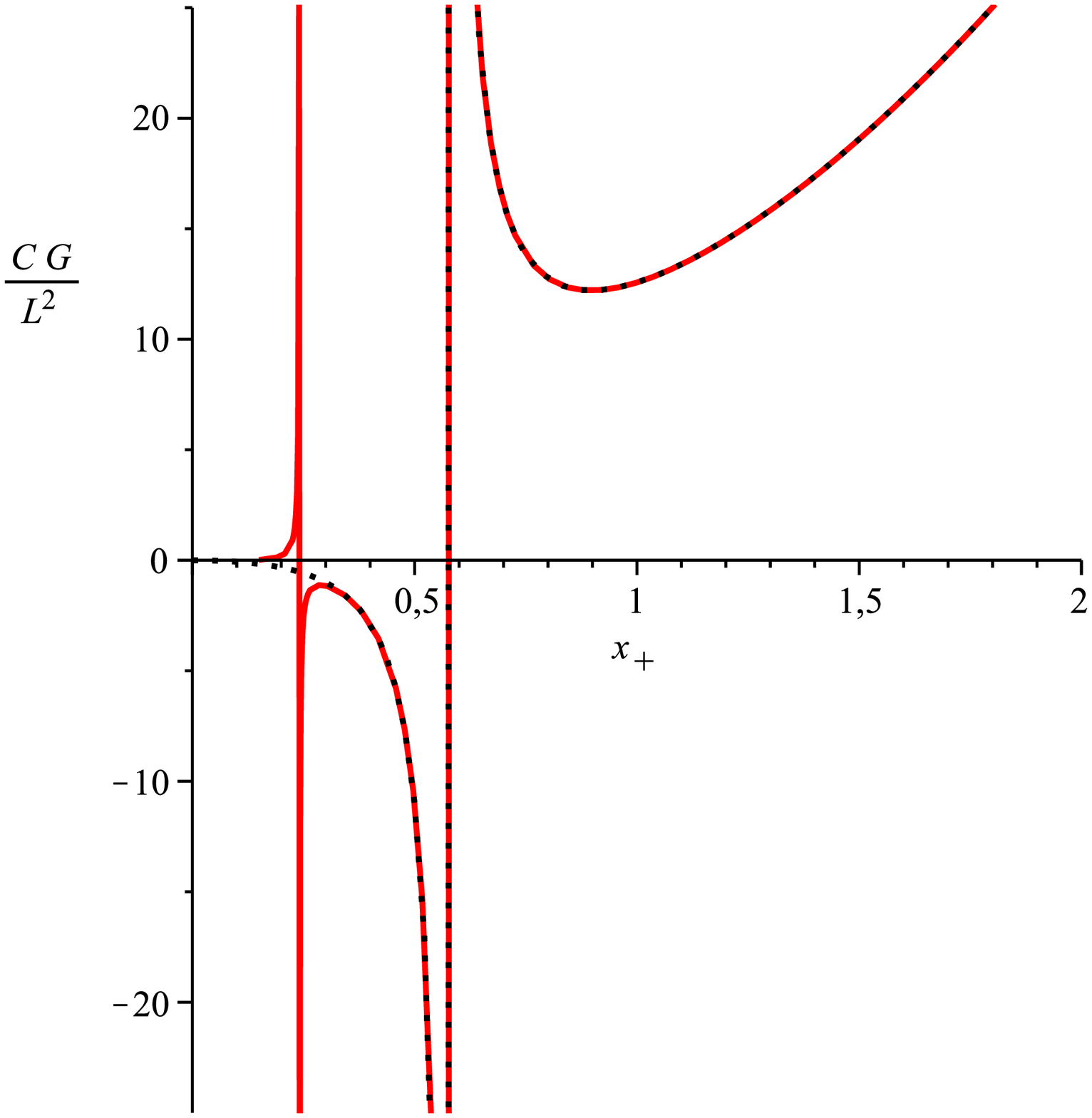}
      \caption{\label{figure2} The dotted curve is the Hawking-Page heat capacity ($q=0$). The solid curves  corresponds to the NCSchwAdS heat capacity for $q=0.1$. We notice that the solid curve is defined only for $x_+\geq x_0\approx 0.15$, which corresponds to the black hole remnant radius. }
     \end{center}
  \end{figure}  
 
Since thermodynamics has a universal character and can be used to derive gravity as an emergent phenomenon, we can  extend the above results to the case in which a specific microscopic structure of the quantum spacetime is prescribed \cite{Pad02}.
As shown in \cite{Nic10}, this can be done by defining a gravitational coupling $\mathbb{G}(r)$ which encodes the short scale quantum gravity modifications with the requirement that at large distances $\mathbb{G}(r)\to G$, i.e., the theory matches Einstein gravity. In the specific case in which the underlying microscopic theory is modelled in terms of noncommutative geometry, we have that the line element (\ref{lineel}) can be cast in the form
\be
V(r)= 1- \frac{2M \mathbb{G}(r) }{r}+\frac{ r^2}{L^2},
\ee
with
\be 
\mathbb{G}(r)=G\frac{2}{\sqrt{\pi}}\gamma(3/2; r^2/4\ell^2).
\ee
We stress that the regularity of the solution is due to the asymptotically safe character of the noncommutative geometry modified gravitational interaction, since at short scales $\mathbb{G}(r)\sim G r^3/\ell^3$.

As a result the internal energy reads
\be 
M=U(r_+)=\frac{r_+}{2\mathbb{G}(r_+)}\left(1+\frac{r_+^2}{L^2}\right)
\ee
and we can compute the new profile of the temperature 
\begin{equation}
T=\frac{1}{4\pi r_+}\left[1+\frac{3r^2_+}{L^2}-r_+\left(1+\frac{r_+^2}{L^2}\right)\frac{\mathbb{G}^\prime(r_+)}{\mathbb{G}(r_+)}\right]
\end{equation}
where
\begin{equation}
\frac{\mathbb{G}^\prime(r_+)}{\mathbb{G}(r_+)}=\frac{1}{4}\frac{r_+^2}{\ell^3}\frac{e^{-r_+^2/4\ell^2}}{\gamma(3/2; r^2_+/4\ell^2)}.
\end{equation}
At large distance the above term is suppressed and one recovers the usual result (\ref{convtemp}). On the other hand at short distances the profile of the temperature is modified: at $r_+=r_0$  the term in square brackets in the above formula admits a zero, corresponding to the case of an extremal black hole configuration ($M=M_0$).
A more accurate understanding of the  temperature behavior is obtained in terms of dimensionless variables. The temperature reads
\be 
T=\frac{L^{-1}}{4\pi x_+}\left[1+3x^2_+ -2q^{-3} x_+^3\left(1+x_+^2\right)\frac{e^{- x_+^2/q^2}}{\gamma(3/2;  x^2_+/q^2)}\right].\nonumber \\
\ee
Plots in Fig. \ref{figure1} disclose the new properties of the solution. In the plot on the left, for  fictitious charges smaller than a critical value $q<q^\ast\approx 0.18243$ the temperature admits both a minimum $T_{\mathrm {min}}$ (as in Hawking-Page case) and a maximum $T_{\mathrm{max}}$, before an evaporation switching off towards a zero temperature extremal black hole remnant configuration. An analysis of the temperature derivative in the plot below shows clearly the presence of two extrema for the temperature. We see that in such regime of the fictitious charge $T_{\mathrm{min}}=\sqrt{3}/(2\pi L)\approx 0.27 L^{-1}$ and $r_{\mathrm{min}}=L / \sqrt{3}\approx 0.57 L$ are essentially unaffected by noncommutative corrections and matches the corresponding values of the SchwAdS geometry. This is not the case for $T_{\mathrm{max}}$ which is sensitive to the values of $q$. This can be explained by the fact that $T_{\mathrm{max}}$ takes place at $r_{\mathrm{max}}<r_{\mathrm{min}}$, i.e., at length scales where the modifications due to $\ell$ are relevant and prevent the occurrence of a divergence in the temperature. The presence of two extrema for the temperature has repercussions on the behavior of the heat capacity $C=dU/dT$
\begin{equation}
C=\frac{\frac{2\pi}{\mathbb{G}(r_+)}r_+^2\ \left[\frac{\mathbb{G}(r_+)}{r_+\mathbb{G}^\prime(r_+)}\left(1+\frac{3r^2_+}{L^2}\right)-\left(1+\frac{r^2_+}{L^2}\right)\right]}{\frac{\mathbb{G}(r_+)}{r_+\mathbb{G}^\prime(r_+)}\left(\frac{3r_+^2}{L^2}-1\right)-\mathbb{H}(r_+)\left(1+\frac{r^2_+}{L^2}\right)-\frac{2r_+^2}{L^2}},
\end{equation}
where
\begin{equation}
\mathbb{H}(r_+)=2-\frac{r_+^2}{2\ell^2}-\frac{r_+\mathbb{G}^\prime(r_+)}{\mathbb{G}(r_+)}.
\end{equation}
Plots in Fig. \ref{figure2} show that the usual vertical asymptote of the Hawking-Page phase transition is accompanied by a second one corresponding to the presence of a maximum  temperature.  This means that the heat capacity switches sign two times, implying an unstable black hole configuration ($C<0$) only for $r_{\mathrm{max}}<r<r_{\mathrm{min}}$. 

\begin{figure}
 \begin{center}
 \includegraphics[height=5.5cm]{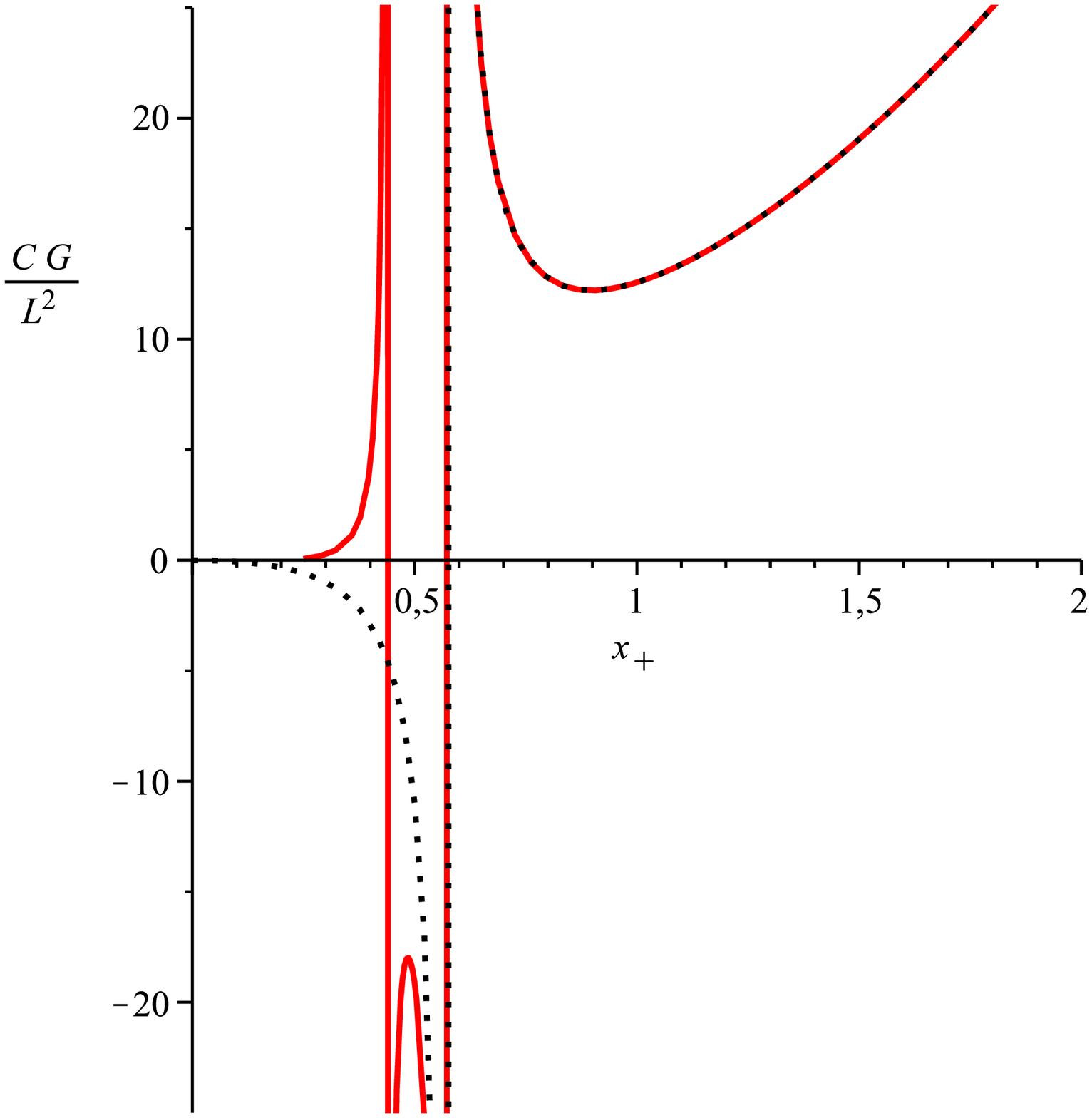}
 \includegraphics[height=5.5cm]{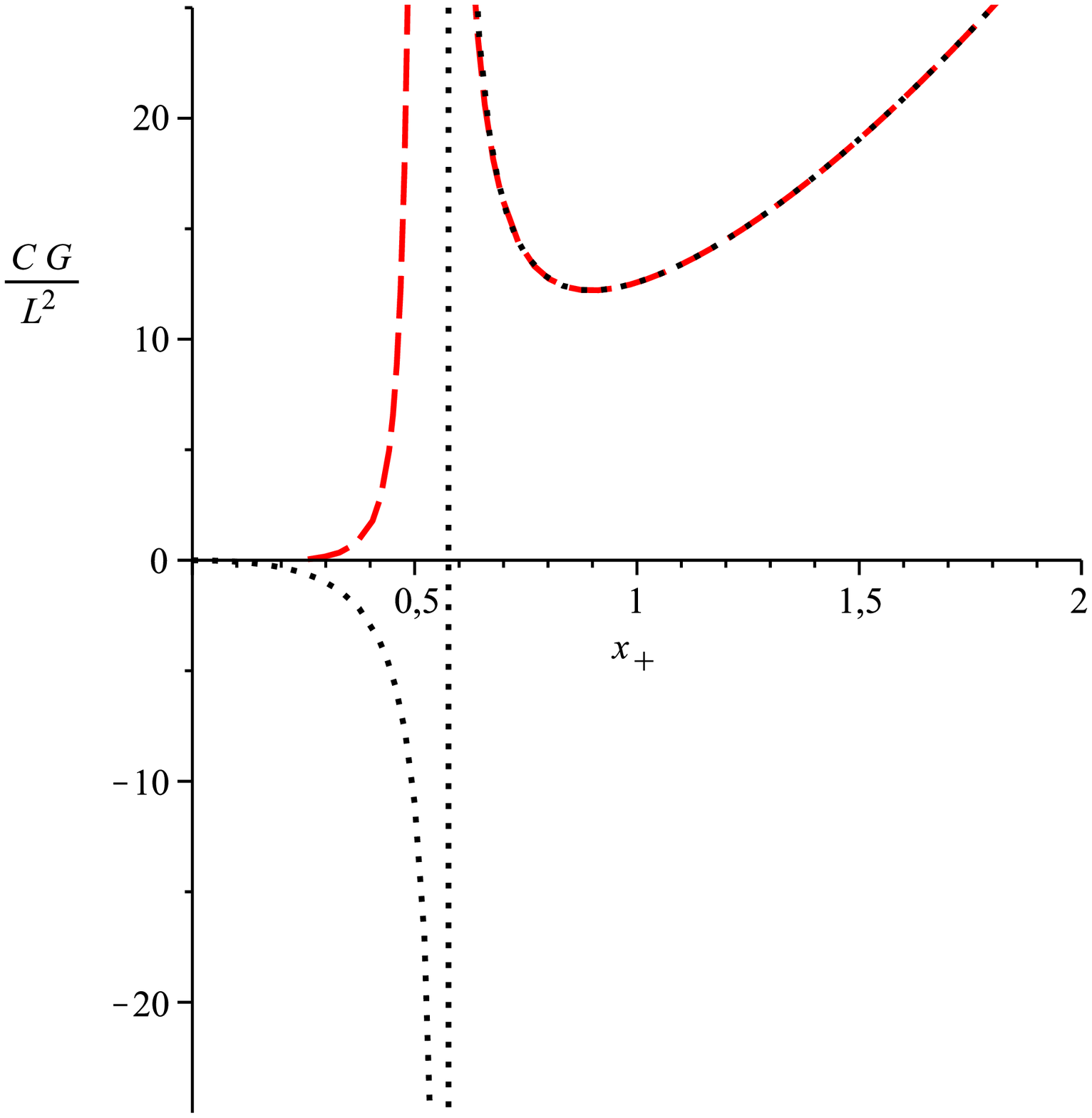}
 \includegraphics[height=5.5cm]{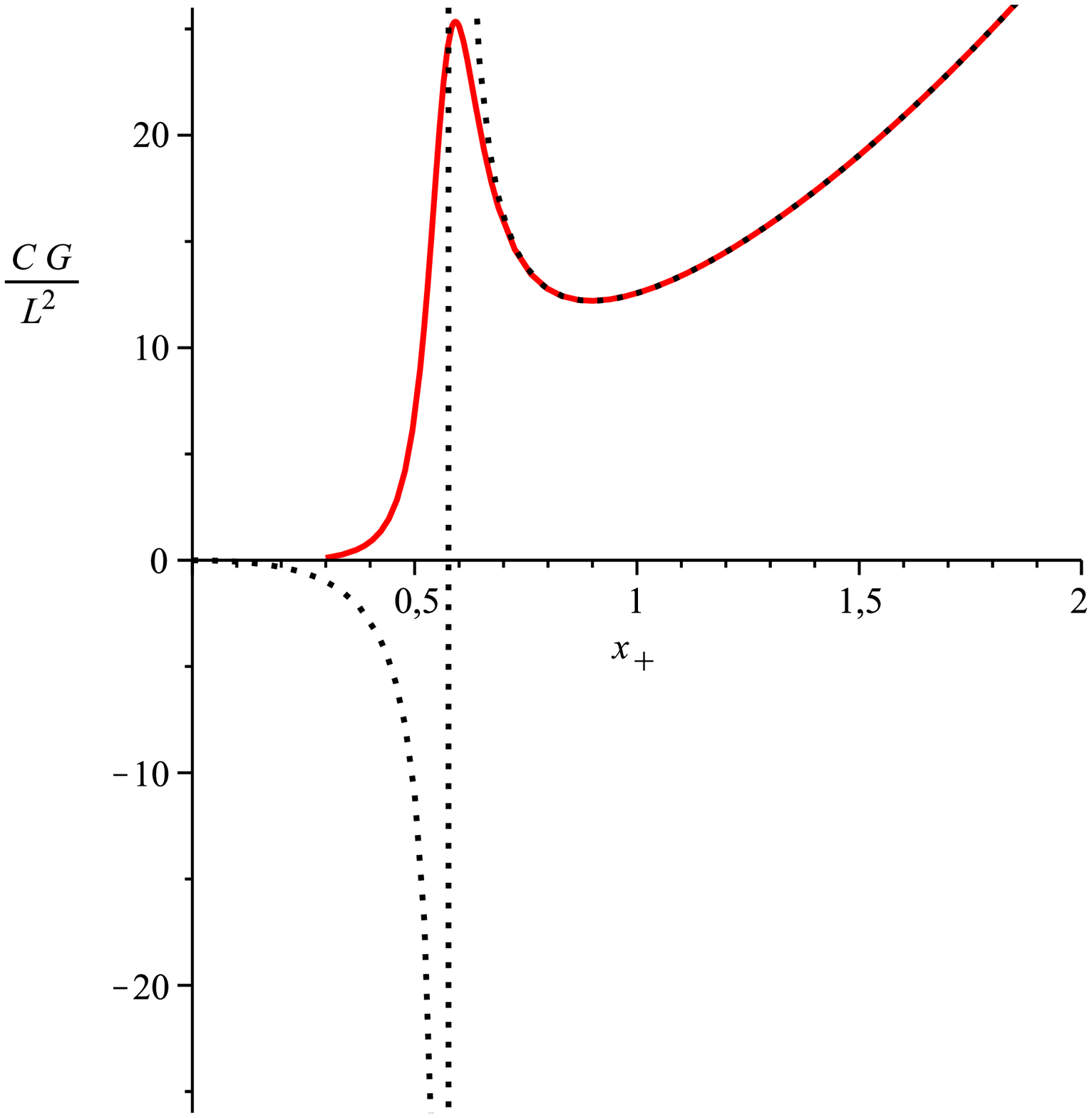}
 \includegraphics[height=5.5cm]{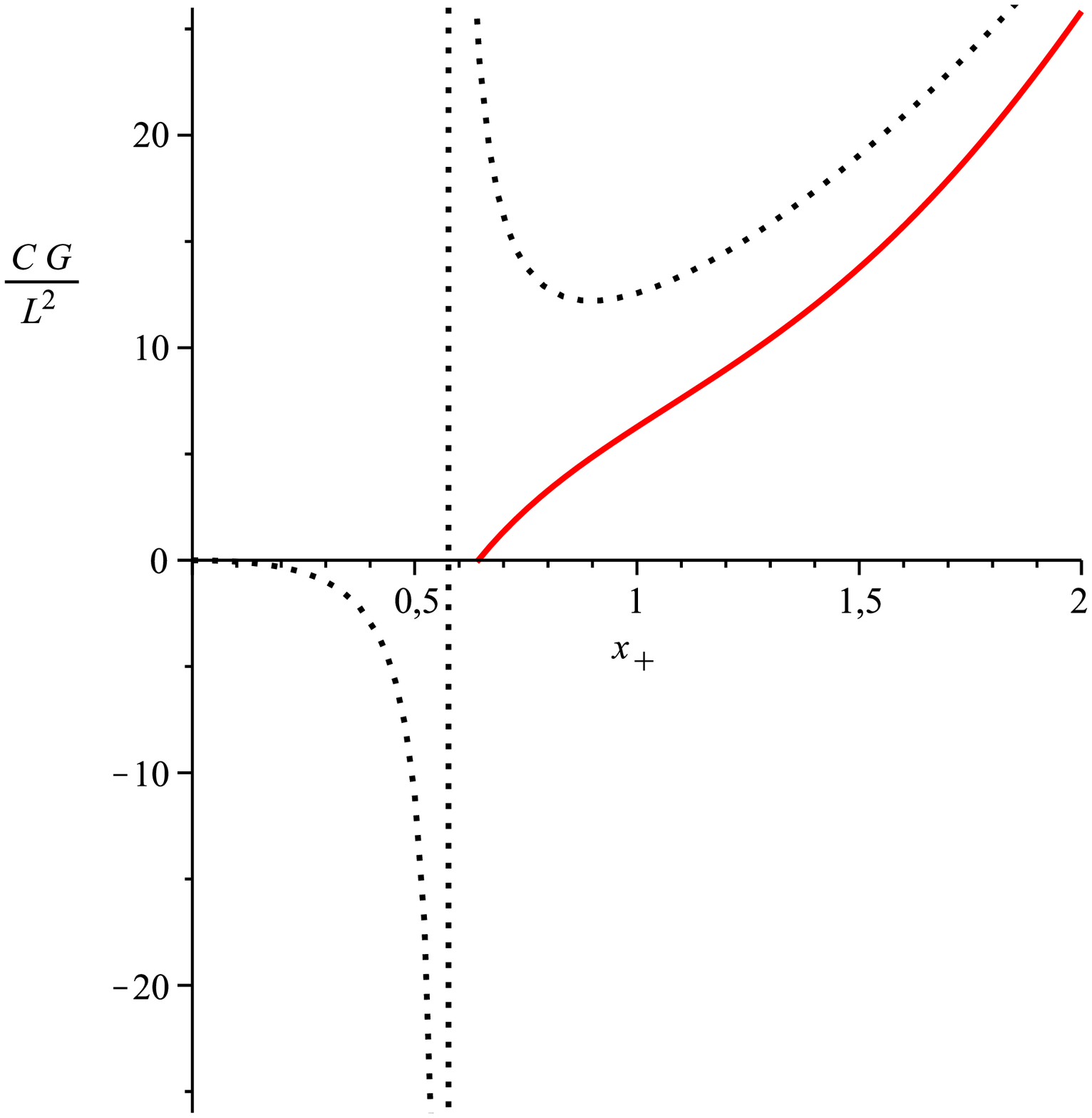}
       \caption{\label{figure3} The dotted curve is the Hawking-Page heat capacity ($q=0$). The solid curves  correspond to the NCSchwAdS heat capacity for $q=1/6$ (top left), $q=0.18243$ (top right), $q=0.2$ (bottom left) and $q=1$ (bottom right). Solid curved are defined for $x_+\geq x_0$, where the remnant radius $x_0$ varies with $q$. }
     \end{center}
  \end{figure}

\begin{figure}
 \begin{center}
  \includegraphics[height=5.5cm]{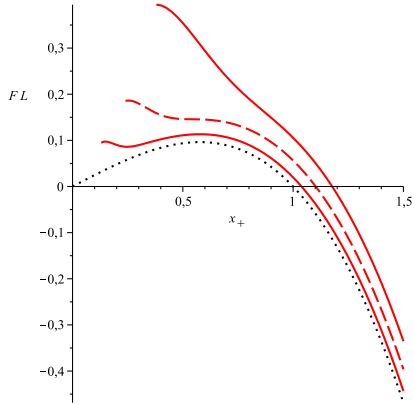}

      \caption{\label{figure4}We have the free energy $F\times L$ for the system in equilibrium as a function of $x_+$ from top to bottom for $q=1/3$ and $q=0.1$ (solid lines). The dashed lines represent the critical case $q=q^\ast$, while the dotted line is the usual SchwAdS case with the Hawking-Page phase transition ($q=0$). 
      }
     \end{center}
  \end{figure}

 For a deeper analysis we need additional thermodynamics quantities.
The entropy $dS=dU/T$ reads
\begin{equation}
S(r_+)=2\pi \int_{r_0}^{r_+} \frac{1}{\mathbb{G}(r_H)} r_H dr_H
\label{entropy1}
\end{equation} 
and the free energy is
\be 
F(r_+)=F_0+ \Delta F(r_+)
\label{newfree}
\ee
where $F_0=U_0$ is the free energy of the extremal zero temperature configuration and 
\be \Delta F=F-F_0= (U-U_0) - TS - S T.\ee
By differentiating the free energy with respect to $r_+$ we get
\be 
&&\frac{\partial F}{\partial r_+}= -S\left(\frac{\partial T}{\partial r_+}\right)\\
&&\frac{\partial^2 F}{\partial r_+^2}= -\frac{C}{T}\left(\frac{\partial^2 T}{\partial r_+^2}\right).
\ee
The plot of the free energy is in Fig. \ref{figure4}.
If we consider the case $q=0.1$ we see that for $T_{\mathrm{min}}<T<T_{\mathrm{max}}$, there exist three possible horizon radii, $r_1$, $r_2$ and $r_3$, with $r_0<r_1<r_{\mathrm{max}}<r_2<r_{\mathrm{min}}<r_3$. 
The largest radius $r_3$ is the less modified with respect to the corresponding radius of the conventional SchwAdS case.
It takes place in a region where the heat capacity is positive defined i.e., at the right of the right asymptote in  Fig. \ref{figure2}. Therefore this configuration is locally stable. However if $r_3<L$, we see from the top plot in Fig. \ref{figure4}, that the free energy is positive and thus the black hole would reduce its free energy by thermal emission as in the conventional case. On the contrary for $r_3>L$, the free energy becomes negative and the black hole configuration turns to be globally stable. For the radius $r_2$ new features start to emerge. As in the conventional case, this corresponds to an unstable configuration ($C<0$ in Fig. \ref{figure4}). On the other hand such configuration has three possibilities: decaying to $r_3$;  decaying to $r_1$; decaying into thermal radiation.
The tunnelling probability for $r_1$ and $r_3$ to occur will be of the form 
\be \Gamma = A e^{-B}\ee
where $A$ is some determinant and $B$ is the difference between the actions of the smaller and the larger radius solutions at the same temperature. Since the instantons associated to each horizon are proportional to $-S$, the black hole entropy,  we have that
\be &&\Gamma_{12} =A e^{S_1-S_2}\\
&& \Gamma_{23}=A e^{S_2-S_3}.
\ee
The tunnelling is favoured towards the larger radius $r_3$ due to the monotonically increasing behaviour of the entropy in (\ref{entropy1}). 
The third option resembles what happens in the conventional case apart its final fate. Instead of a complete evaporation into pure thermal radiation, the black hole, due to the negative heat capacity, keeps radiating at increasing rate until it reached the maximum temperature. Then the free energy reaches a local minimum and the heat capacity admits a vertical asymptote before switching to a locally stable configuration ($C>0$). 
The smallest radius $r_1$ is a completely new case. It occurs in the region where the heat capacity is positive defined, just on the left of the left asymptote (see Fig. \ref{figure2}). However it is a globally unstable configuration since the free energy is positive. The black hole would reduce its free energy if the black hole evaporated completely. On the other hand, by emitting radiation the black hole reduces its radius, its temperature and its evaporation rate.  At $r_0$ the temperature and the heat capacity vanish and the evaporation stops. The free energy does not reach negative values but admits a local maximum. This amount of free energy in $r_0$ can be explained with the energy stored in the extremal configuration, i.e, $F=U_0=M_0$.
We recall the our model of quantum spacetime is endowed with an effective lattice whose quantum cells have size $\ell$. The absence of resolution at scales smaller than $\ell$ implies a minimum mass $M_0$. We notice that in this regime of $q$, the remnant extension is about just few of these quantum cells. For $q=0.1$ we find $r_0\approx 0.15 L=3\ell$.

By increasing the fictitious charge $q$, the two asymptotes get closer, corresponding to a decreasing of the unstable configuration $r_{\mathrm{max}}<r_+<r_{\mathrm{min}}$ (see the top plot in Fig. \ref{figure3}). Eventually the two asymptotes coalesce, i.e., $r \ast=r_{\mathrm{max}}=r_{\mathrm{min}}$ for the critical value of the fictitious charge $q=q^\ast\approx 0.18243$ (see the top right plot in Fig. \ref{figure3}). In such a case the temperature becomes a monotonically non decreasing function of $r_+$ and we enjoy the following properties
\be
&&\frac{\partial T}{\partial r_+}(r^\ast) = \frac{\partial^2 T}{\partial r_+^2}(r^\ast)=0\\
&& \frac{\partial F}{\partial r_+}(r^\ast) = \frac{\partial^2 F}{\partial r_+^2}(r^\ast)=0.
\ee
The heat capacity turns out to be positive defined. The presence of the asymptote at $r_+=r^\ast$ is a sign of a phase transition between large ($r>r^\ast$) to small ($r<r^\ast$) black holes. Both black hole configurations are locally stable, even if globally unstable (the free energy is positive unless $r_+>1.1L$). By evaporating the black hole cannot decrease its free energy (see plot at the top in Fig. \ref{figure4}), which reaches a maximum in the extremal configuration. This means that lower mass black holes would be less probable than the larger mass ones.

For higher values of the fictitious charge, i.e., $q>q^\ast$, the temperature is a monotonically increasing function of $r_+$. The heat capacity is positive defined and depending on the values of $q$ can or cannot admit a local maximum, but never diverges. As a result no phase transition occurs in such a regime. For an inspection of the behaviour of the free energy, the smaller mass black holes would be less probable with respect to the larger ones  (see plot at the top in Fig. \ref{figure4}),  which turns out to be globally stable for $r_+>1.2L$.

\subsection{Classification of phase transitions}

Phase transitions are classified by the analytical structure of the parameter characterizing them, the so called ``order parameter''  (in the Hawking-Page transition this is the Schwarzshild radius).  Phase transitions are labelled by the order of by the lowest derivative of the free energy with respect to the order parameter that is discontinuous at the transition. As a result transitions with a discontinuity in the order parameter are called first order, transitions with a discontinuity in its first differential are called second order.  In an equivalent but maybe more modern perspective phase transitions are broadly classified in two categories:  transitions with  latent heat (i.e. fist order phase transitions) and transitions without latent heat (i.e. continuous phase transition).  First order phase transitions involve a latent heat because the entropy (i.e. the first differential of the free energy) shows a discontinuity. Continuous phase transitions include real phase (transitions with a discontinuity of the order parameter at a given order) and analytic crossovers, i.e., transitions where the order parameter and all its differentials are continuous.

This classification can be understood \cite{huang,BMS10} in terms of the minima of the free energy:   if the free energy with respect to the order parameter has two coexisting local minima, the phase transition will be first order (as the free energy smoothly changes with temperature, one minimum can become smaller than the other, i.e., this is a discontinuity).  If degenerate minima merge, it signals a second order transition, while if the structure of minima is unchanged it usually means a crossover.   Because the position of free energy minima is related to the system's symmetry, a spontaneous breaking and restoration of symmetry signals a phase transition (the converse however is not the case: plenty of phase transitions, for example the van der Waals one described below are {\em not} associated with a change in symmetry).

 Sometimes a crossover can become a first order phase transition.  The point at which it happens is called a critical point, and the dynamics in its vicinity is identical to the dynamics in the vicinity of a second order transition.  A feature of this dynamics is ``universality'', insensitivity of the scaled value of the order parameter to microscopic dynamics.  Thus microscopically very different systems will have a very similar dynamics close to a critical point.  

\begin{figure}
 \begin{center}
   \includegraphics[height=5.5cm]{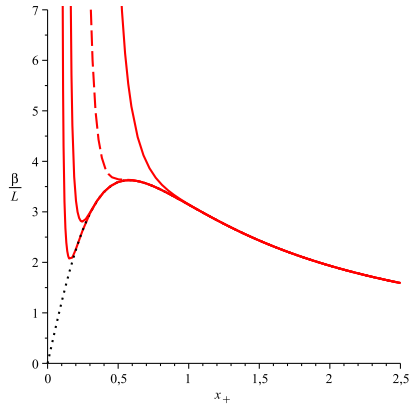}

      \caption{\label{figure5}The function $\beta(r_+)=T^{-1}$ from top to bottom for $q=1/3$, $q=0.1$ and $q=1/15$. The dashed lines represent the critical case $q=q^\ast$, while the dotted line is the usual SchwAdS case with the Hawking-Page phase transition ($q=0$).   
      }
     \end{center}
  \end{figure}

The phase structure of the NCSchwAdS solution is plotted in Fig. \ref{figure5}. The reader will recognize the plot as the classic phase diagram of the liquid-gas system within the van der Waals theory. To translate, the fictitious charge $q$ has to be interpreted as the temperature of the fluid, $\beta$ as the pressure and $r_+$ as the volume. The van der Waals gas is the simplest model which includes a non-zero size of molecules, preventing the gas to move in all the volute of the container because some of the volume is occupied by other gas molecules. It is not hard to understand that our model of quantum spacetime, by introducing a minimal length $\ell$, has the same role with respect to the conventional SchwAdS spacetime as the van der Waals gas has with respect to an ideal gas. In the isomorphism between our parameters and those of the van der Waals system, our remnant radius $r_0\sim\ell$ is equivalent to $b$, the parameter accounting for the volume excluded owing to the finite size of molecules. If $q$ is set to zero, then $b$ is vanishing and one recovers the equation of state of an ideal gas (SchwAdS system). Moreover in the low density limit, $r_+\gg r_0$, we also recover the ideal gas behavior. When the density is high and $r_+$ approaches $r_0$, the ``pressure'' $\beta$ shoots up. In the language of black holes, this is equivalent to say that the ultra-high temperature phases are dominated by non-extreme black holes while zero temperature remnants may form at short scales.
 
As the ``temperature'' $q$ is lowered the isotherms change from being somewhat ideal-gas type to exhibiting an S-type shape with a minimum and a maximum. This fact has consequences on the gas ``compressibility'' $\kappa\equiv - \frac{1}{r_+}(\partial r_+/\partial \beta)$, which admits a region of negative values. This is a situation for which a gas compression implies a gas expansion. In the language of black holes this is equivalent to say that an increase of the temperature results in a  decrease in the black hole radius and hence mass ($C<0$). This cannot a be a stable situation: if a ``pressure'' fluctuation temporarily increases the ``pressure'' $\beta$, the ``volume'' $r_+$ increases and a negative work is done on the gas, providing energy to increase the pressure fluctuation. Thus in this parameter region, the system is unstable with respect to fluctuations. This problem occurs when the ``temperature'' $q$ is lower than the critical value $q^\ast$. At the critical ``temperature'' $q^\ast$ the curve does not have a maximum or minimum, but shows a point of inflection, known as the \textit{critical point}.

The analogy between the two systems lets us explore further features: for ``temperatures'' $q<q^\ast$ the free energy becomes a multiply valued function for certain values of the ``pressure'' $\beta$. Since the ``pressure'' $\beta$ will minimize the free energy, the system will actually assume a single (the lowest) value for the free energy. However at each $\beta$, there exist two corresponding values of $r_+$ in the locally stable branches, 
 i.e., $\partial r_+/\partial \beta<0$. By the Maxwell construction, we have that phases corresponding to these two values of $r_+$ can be in equilibrium with each other. Therefore we have a two-phase region in which gas (large black holes) and liquid (small black holes) can coexist. From the plot in Fig. \ref{figure4}, we see that the free energy of small black holes is positive while that of large black holes is negative, so that the branch of non-extremal black holes dominates. This is the situation of a genuine phase transition of the first order, which result in a jump from the stable branch of small black holes to that of large black holes. Being the entropy proportional to $r_+^2$, there is a jump in the entropy, or release of latent heat.
 
 At the ``critical temperature'' $q=q^\ast$, the unstable branch disappear. Thus the transition between small and big black holes occurs without latent heat, corresponding to the case of a continuous phase transition with singular heat capacity.
 
 For ``temperatures'' $q>q^\ast$, the free energy and the ``volume'' $r_+$ are single valued functions and there is simply a crossover between a system with low compressibility (small black holes) to one with progressively higher compressibility (large black holes) as the ``pressure'' $\beta$ is reduced.

\section{Relevance for gauge-string duality?}
\label{gauge}

One of the most interesting and promising theoretical achievements in the 
last few decades has been the gauge-gravity correspondence, a conjectured 
duality relating $d$-dimensional gravitational physics to $d-1$ 
dimensional quantum field theory 
\cite{ads1,adswitten,adswitten2,ads2,ads3}.
The most well-studied special case of this correspondence is the so-called AdS/CFT \cite{ads1}, relating Anti-de sitter space to $N=4$ SYM theory, which is thought to be conformally invariant in four dimensions.   Subsequent work, however, has shown the duality is more general than this, with boundaries breaking some of the profound theories of CFT dual to complicated brane set-ups in asymptotic AdS.
For example, $D7$ branes effectively add a global flavor symmetry to the theory \cite{mateos}, while conical membrane set-ups \cite{mohammed,kstrassler} and dilaton potentials \cite{gubser,kiritsis,jorge} can be used to mimic a non-conformal $\beta$-function.


One of the original pieces of evidence for AdS/CFT was the relationship, explored in \cite{adswitten,adswitten2} of the Hawking-Page phase transition to deconfinement, in the specific case geometrical deconfinement on spherically compactified spaces.    Thus, do the results presented in this work have any relevance to gauge-string duality?
While we can not rigorously prove it, we conjecture that indeed this is the case.

The first thing to remember is that, while the $AdS_3$ case presented here is considerably different from the $AdS_1 \times S_n$ examined in \cite{adswitten,adswitten2} or the $AdS_5 \times S_5$ background examined in \cite{ads1}, our results fall neatly into the Van Der Waals universality class.
Because of that, and the structure of the calculations in the previous section, we expect that the general structure of this phase diagram will be maintained in $AdS_1 \times S_n$ or any other background where a deconfinement phase transition exists.
The critical $q^*$ will of course be different, but not the behavior around the critical point.       This critical point will translate itself in the behaviour of $N_c$ of the boundary theory, once again in a way that is universal around the critical point but sensitive to the exact nature of the background theory.

When the number of colors $N_c \gg$ the number of flavors $N_f$ (the so-called T'Hooft limit), confinement is rigorously known to be a phase transition in all dimensions greater than one, since the symmetries of the system are expected to change. The low temperature phase will have acquired an extra $Z_N$ symmetry, signifying a zero expectation value of the Polyakov loop \cite{poldef1,poldef2}.
In the high-temperature phase the Polyakov loop will aquire a finite expectation value, and $Z_N$ will be spontaneusly broken.

At finite number of flavors $N_f$, the $Z_N$ symmetry is no longer exact at a fundamental level.  Hence, it is natural to expect that at some critical $N_c^{crit}$ ``a critical point'' for deconfinement appears, where for higher $N_c$ deconfinement is a phase transition and for lower $N_c$ it is a crossover.  This is indeed what emerges from lattice simulations \cite{pollat1,pollat2} in three dimensions with a finite number of flavors.   The free energy with respect to the Polyakov loop expectation value $\ave{L}$ and energy $E$ is therefore of the form
\begin{equation}
F\left( \ave{L},E \right) = F_{gauge} \left( \ave{L},E \right) +  \delta F \left(\frac{N_f}{N_c}, \ave{L},E \right).
\end{equation}
$F_{gauge}$ at the mixed phase is a function with two local minima \cite{poldef2}, corresponding to the two phases.  For high-enough $N_f/N_c$, however, $\delta F$ could destroy one of the minima and make the phase transition into a crossover.  Substituting $E$ for the black hole mass and $\ave{L}$ for the horizon $r_+$, this could correspond to Eq. \ref{newfree}.  Note that while $\delta F$ goes to zero when $N_f/N_c$ go to zero, it is {\em not} guaranteed this happens in a way that can be Taylor-expanded, and indeed in Eq. \ref{newfree} this is not the case (strictly speaking, the critical point might only be reached at $N_f=0$, ala Gross-Witten \cite{gwpoint}. If our result is correct, this {\em is not the case}, and lattice seems {\em not} to support this scenario)

In the gauge-gravity correspondence,
 changing $N_c$ at fixed coupling constant $\lambda$ is equivalent to changing the string coupling constant $g_s$ while mantaining the worldsheet coupling constant $\alpha'$ the same.    Hence, modeling the contribution of $\delta F$, and how it may change the order of the phase transition, is inherently a problem that goes beyond classical gravity.   

As we saw, on the field theory side, the absence of a crossover at $N_f/N_c=0,\Delta F=0$ and its appearance at finite $N_f/N_c,\Delta F$ is dictated by symmetry principles.
On the string theory side, the order of the Hawking-Page phase transition is similarly determined by widely believed results about classical gravity and their breaking in the quantum-gravity regime.
The fact that Hawking-Page is a phase transition rather than a cross-over is a restatement of the widely-believed cosmic censorship hypothesys \cite{penrose}, the idea that any singularity in spacetime is surrounded by an event horizon (the black hole's existance therefore presupposes a discontinuity in energy-momentum, i.e., a phase transition).   Hence, to break cosmic censorship is equivalent to smoothening the singularity, something that requires quantum corrections, parametrized in our ansatz by a size $q \times$ the macroscopic size of the system under consideration.

Saying something more quantitative rigorously at the moment is very difficult, 
since string loops can not be yet calculated  in any of the backgrounds thought to be dual to gauge theory.   The fact that we are dealing with a critical point on both sides of the correspondence means that we might be able to learn something from universality-based arguments without understanding the full microscopic dynamics of the system.  The noncommutative geometry model described in this work successfully describes the critical point on the Hawking-Page side, and hence we can use it as an ``effective description'' of a finite $N_c$ theory in 2+1 dimensions.

Here, we conjecture that $q$ measures $N_f/N_c$.   
Qualitatively, the behavior seen here seems to be correct. At $q=0$ you always have a phase transition.   For small $q$'s, the phase transition continues, but becomes a cross-over for larger $q$'s.
In the AdS world, $N_f$ is typically rapresented by branes.    The regime in which $N_f$ is non-negligible with respect to $N_c$ should correspond to the regime where backreaction of branes by strings is not-negligible.   While {\em some } aspects of this back-reaction show up at the classical level, $g_s\ll 0$ corresponds to the limit in which the microscopic ``size'' of the brane approaches the string size.      The usual arguments about noncommutative geometry appearing as an effective description of quantum gravity therefore become applicable \cite{Nic09}, and, as shown for example in \cite{gravbrane} it is possible to obtain noncommutative gravity of the sort studied here by stringy corrections.

Our conjecture has some rather unconventional, but very reasonable and testable, prediction.  The phenomenological significance of large $N_c$ calculations \cite{largenc1,largenc2} depends on the large $N_c$ being reached smoothly. 
Little indication, however, exists that this is the case.
In fact, from the beginning of this field, the Skyrme crystal picture of nuclear matter in \cite{skyrmecrystal} it seemed quite likely that a {\em phase transition} in $N_c$ space was likely \cite{ncphase1,ncphase2,ncphase3} at least in dense nuclear matter (\cite{skyrmecrystal} points out that baryon mass and interaction strength grows with $N_c$, and posits that at a certain crytical $N_c$ the ``Skyrme Crystal'' baryonic state derived there melts into the loosely-bound nuclear liquid we are familiar with.  Since this implies a breaking of crystal symmetry, this must be a phase transition rather than a crossover).
Hence, extrapolating from $N_c$ large to $N_c=3$ is not just a very rough approximation, but an untenable one, as in-between there might be a discontinuity.

Recently the claim has been made that this transition is percolation-driven \cite{ncphase1}.   Essentially, when $N_c$ is varied at constant baryonic density, quarks of different baryons introduce long-term correlations at critical $N_c$.  Physically, this could be related to the violation of cosmic censorship due to quantum-gravitational effects. The  critical point manifests itself as a second order phase phase transition, 
and percolation is a second order transition in any dimension greater than one.
Moreover, the Hawking-Page transition coincides with the transition from a 
gas of black holes to one large black hole.
It is plausible that  the ``crossover" regime occurs because the  
black holes in the gas are actually  
linked by super-horizon quantum gravitational interactions, modeled in our approach by ``smearing'' the super-horizon black hole distance by a Planck-sized width.  
Such super-horizon links look qualitatively very similar to the percolation links in 
\cite{ncphase1}.   

The analogy with the van der Waals gas, described in the previous section, also finds a natural place within our conjecture.  On the string side, the critical point and crossover are triggered by the interplay between macroscopic scales (cosmological constant and black hole size) and the quantum ``size'' of spacetime (or, if one wants, of the string vacuum loop).
On the gauge side, when $N_c$ is varied in the dense matter strongly coupled regime, deviations from the ``classical'' skyrme crystal arise due to the interplay between the number of colors and the number of neighbours $N_N$ in the densely packed system \cite{ncphase3}.  This interplay, unavoidable due to Pauli's exclusion principle, separates the $N_c/N_f \gg N_N$ phase, effectively classical and strongly bound \cite{skyrmecrystal} from the $N_c/N_f \ll N_N$ phase (relevant to our world), where quantum effects weaken the bound between baryons and, presumably, melt the skyrme crystal into the nuclear gas and transform the deconfinement phase transition into a crossover.
   The correspondence with a van der Waals-like Hawking-Page phase diagram is manifest.
As percolation is a geometrical transition, the picture in terms of $N_N$ complements the percolation picture described in the previous paragraph.

If our speculations are correct, $q^{*}=0.18243$ should be related to the critical number of $ N_f/N_c$ required for percolation in 2D \cite{ncphase1} and to the critical number of colors at which Pauli-blocking with neighbouring baryons become important \cite{ncphase3}.  The theoretical uncertainity regarding the relationship of our model to string theory is too large for the $\order{1}$ factor of this relation to be calculated explicitly, but assuming it is unity, the critical number of colors for one flavor is 6, a result remarkably close to geometrical and percolation arguments seen in \cite{ncphase1,ncphase3}.     We are therefore confident that the speculations here have some merit, although of course a full lattice (2+1D for the current work, 3+1D for the forthcoming 5D Hawking-Page extension) calculation on the numerical side, and a better understanding of how $q^*$ is related to $N_c$ at fixed $\lambda$ ($\sim \order{5-30}$ from phenomenological constraints \cite{adspheno2}) on the analytical side, is necessary.

\section{Conclusions \label{concl}}

In this paper, we have derived the noncommutative geometry inspired Schwarzschild anti-deSitter solution which models quantum gravity effects we expect in place of the classical curvature singularity. We have presented an extensive analysis of the solution both on the geometric and thermodynamic side. We found that the geometry admits two, one or no horizon depending on the value of the mass parameter with respect to a threshold mass $M_0$ set by the noncommutative geometry induced minimal length $\ell$ and the cosmological parameter $L$. In agreement with the tenets of quantum gravity, our effective geometry is singularity free. In addition the possibility of horizon extremisation has repercussions on the thermodynamics of the solution: instead of a divergent behavior at short scales, the black hole temperature admits a maximum before a cooling down towards a zero temperature remnant configuration. As a consequence of this new profile of the temperature, also the phase structure of the solution is modified: 
 we have shown that 
 the Hawking-Page phase transition extends into a phase diagram with a critical point. When the quantum-gravitational scale, parametrized here by $q=2\ell/L$, is above a critical value $q^*$, the Hawking-Page transition becomes a crossover with no singular behaviour in any thermodynamic quantities.
We have speculated that in the context of gauge-gravity duality, this corrisponds to the
 critical point in $N_f/N_c$ space.   If this is the case, reproducing the noncommutative limit with string theory techniques, and measuring the critical $N_f/N_f$ on the lattice would be the first real piece of evidence that the gauge-gravity correspondence is valid beyond the classical gravity limit.

\begin{acknowledgments}
This work is supported by the Helmholtz International Center for FAIR within the
framework of the LOEWE program (Landesoffensive zur Entwicklung Wissenschaftlich-\"{O}konomischer
Exzellenz) launched by the State of Hesse.  We would like to thank Rene Meyer for critical discussions and suggestions.

\end{acknowledgments}
\providecommand{\href}[2]{#2}\begingroup\raggedright\endgroup

    \end{document}